\documentclass[default]{svmult}

\usepackage{makeidx}         
\usepackage{graphicx}        
\usepackage{caption}
\usepackage{subcaption}
\usepackage[bottom]{footmisc}

\usepackage{newtxtext}       %
\usepackage[varvw]{newtxmath}       

\usepackage[section]{placeins}

\makeindex             

\begin{document}

\title*{Cats vs Dogs, Photons vs Hadrons}
\author{Francesco Visconti}
\institute{Francesco Visconti \at Osservatorio Astronomico di Roma, INAF, via Frascati 33, Monte Porzio Catone (RM), \email{francesco.visconti@inaf.it}}

\maketitle
\vspace{-0.5in} 

\abstract{In gamma ray astronomy with Cherenkov telescopes,  machine learning models are needed to guess what kind of particles generated the detected light, and their energies and directions. The focus in this work is on the classification task, training a simple convolutional neural network suitable for binary classification (as it could be a \emph{cats vs dogs} classification problem), using as input uncleaned images generated by Montecarlo data for a single \textbf{ASTRI} telescope. Results show an enhanced discriminant power with respect to classical random forest methods.}

\section{The context: detect Cherenkov light with ASTRI telescopes}
\label{sec:context}
Ground-based gamma-ray astronomy is a young field with enormous scientific potential. The current generation instruments \emph{H.E.S.S.}\footnote{\url{https://www.mpi-hd.mpg.de/hfm/HESS/HESS.shtml}}, \emph{MAGIC}\footnote{\url{https://magic.mpp.mpg.de/}} and \emph{VERITAS}\footnote{\url{https://veritas.sao.arizona.edu/}} have already demonstrated the huge physical potential of astrophysical measurements at teraElectronvolt (TeV) energies. The 250+ sources detected by these instruments (\url{http://tevcat.uchicago.edu/}), and the wide range of high impact scientific results, suggest that particle acceleration yielding TeV gamma-rays is common in nature.

The \emph{ASTRI}~\cite{Scuderi22} (Astrofisica con Specchi a Tecnologia Replicante Italiana) telescope is an INAF project falling in the category of Cherenkov Telescope Array's Small Size Telescope (SST) with a diameter of 4 m, with double mirror technology. The ASTRI's data reduction software \cite{Lombardi18} has been entirely developed by a dedicated INAF group of work. This work is focused on the reconstruction stage, and in particular in the piece of software needed to discriminate between real signal and background.

\section{Network and Data}
\label{sec:network}
Any data analysis software for Cherenkov telescopes needs a fundamental component to discriminate if the air shower has been produced by photons or cosmic rays; this is done with machine learning techniques, training models on simulated data. 

\begin{figure}[!h]
 \begin{subfigure}[b]{0.5\textwidth}
    \includegraphics[width=\textwidth]{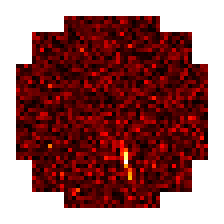}
    \caption{gamma}
    \label{fig:p1}
  \end{subfigure}
  \hfill
  \begin{subfigure}[b]{0.5\textwidth}
    \includegraphics[width=\textwidth]{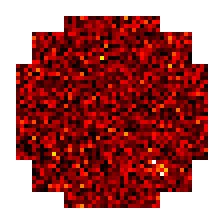}
    \caption{proton}
    \label{fig:p2}
  \end{subfigure}
  \caption{An example of calibrated Cherenkov events as seen by ASTRI telescope.}
\end{figure}

The gold standard in this field is Random Forest~\cite{Breiman}, here implemented with \verb|scikit-learn| library \cite{sklearn} which is run over Hillas parameters: this method introduces a bias (Hillas parameters are calculated after image cleaning and assuming a shape of the Cherenkov light in the image), which can be overcome by deep learning techniques \cite{Shilon}, here implemented with the \verb|keras|~\cite{keras} library.

With that aim, a five layers Convolutional Neural Network has been trained with ASTRI calibrated images from a Montecarlo production. 
For input data, single telescope images have been generated starting from a Montecarlo ASTRI data production, calibrated via the dedicated software developed in Rome by our group~\cite{Lombardi18}. Events data are selected before the cleaning process, keeping the full background of the pixels due to the Night Sky Background (NSB). Those images are the signal received on any given camera pixel, a way different input from Hillas parameters (tabular data) characterizing the events. 70k images have been used for each class in the training step, 30k for validation during training, and finally prediction are made on 100k files for each class.

In Table~\ref{tab:1}, the network architecture adopted for the task is shown.
\begin{table}[ht]
\begin{center}
\footnotesize
\begin {tabular}{ccc}
\hline
Layer (type) & Output Shape & n. of Params \\
\hline
\hline
Conv2D & (None, 54, 54, 32) & 896 \\
\hline
MaxPooling2D & (None, 27, 27, 32) & 0 \\
\hline
Conv2D & (None, 25, 25, 32) & 9248 \\
\hline
MaxPooling2D & (None, 12, 12, 32) & 0 \\
\hline
Conv2D & (None, 10, 10, 64) & 18496 \\
\hline
MaxPooling2D & (None, 5, 5, 64) & 0 \\
\hline
Dense & (None, 64) & 102464 \\
\hline
Dropout & (None, 64) & 0 \\
\hline
Dense & (None, 1) & 65
\\
\hline
\hline
\\
Total (all trainable) params: 131,169 \\
\end{tabular}
\end{center}
\caption{
Convolutional Neural Network architecture used for the classification task.
}
\label{tab:1}
\end{table}

\section{Results}
\label{sec:results}
The training took 1 hour on a K20 NVIDIA GPU, stopped after 11 epochs monitoring the validation accuracy improvement. With regard to the performance, the max validation accuracy reached was 96.7\%. For a quick comparison, the standard method employed in the ASTRI Mini Array data analysis and reconstruction (Random Forest on Hillas parameters) do not exceed 86\% in our case.
The good result is confirmed by the gamma/hadron separation plot\footnote{This kind of plot is an histogram where for each probability bin we count (actually there's a density on the \emph{y} axis) the number of particles classified as gamma with that probability: the more the two populations are separated, the better}, here plotted in log-scale to better appreciate the distributions, where we can see an enhanced discrimination power for the CNN network on calibrated data.
All other metrics taken into account (ROC curve with AUC score, Precision vs Recall, Confusion Matrix) confirm that \textbf{CNN performs signiﬁcantly better than RF}.

\begin{figure}[!h]
    \centering
    \includegraphics[width=0.9\textwidth]{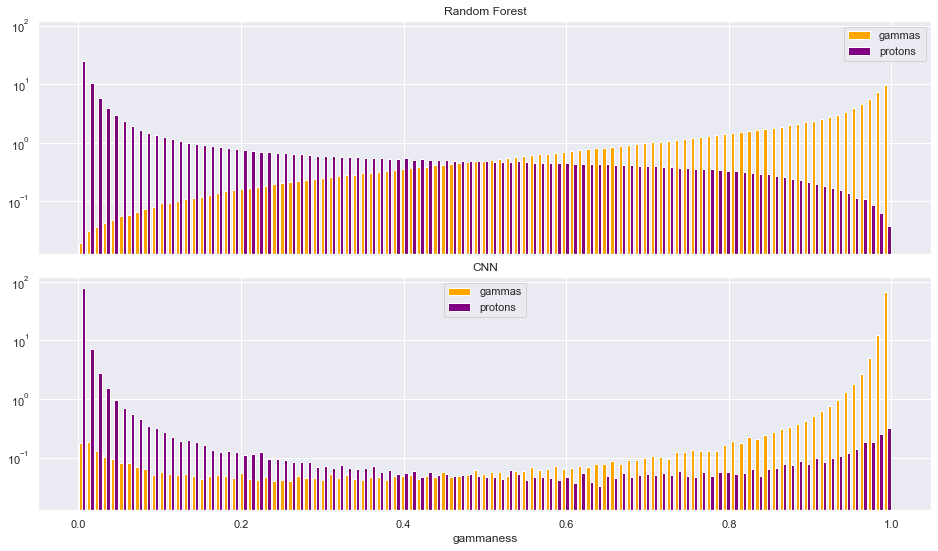}
    \caption{gamma - hadron separation. Above Random Forest, below CNN.}
    \label{fig:gh_sep}
\end{figure}

In Table~\ref{tab:2}, other useful metrics\footnote{For a list of useful metrics see for example https://scikit-learn.org/stable/modules/model\_evaluation.html} are summarized, all pointing in the same direction: this CNN outperforms the Random Forest.

\begin{table}[ht]
\begin{center}
\footnotesize
\begin {tabular}{ccccccccc}
\hline
& Val Acc & AUC & f1 & brier loss & TP & FN & FP & TN \\
\hline
\hline
Random Forest & 86\% & 0.94 & 0.92 & 0.09 & 84\% & 16\% & 11\% & 89\%\\
\hline
Convolutional NN & \textbf{96.7\%} & \textbf{0.99} & \textbf{0.97} & \textbf{0.03} & \textbf{96\%} & \textbf{4\%} & \textbf{3\%} & \textbf{97\%}\\
\hline
\end{tabular}
\end{center}
\caption{
Classification metrics comparison between Random Forest and CNN.
}
\label{tab:2}
\end{table}

\section{Summary}
\label{sec:summary}
The work produced interesting results with a small neural network, enhancing the discriminatory power for the gamma-hadron separation problem. This encourages to explore these techniques, suggesting the bias introduced by event parametrization is very important. Future improvements are currently in progress using "stereo" images (events seen by more than one telescope), with a pipeline able not only to do signal-background classification but also energy and direction reconstruction \cite{Miener}.

\end{document}